\documentclass[floatfix,showpacs,superscriptaddress,nofootinbib,onecolumn,showkeys]{revtex4-2}

\usepackage[utf8]{inputenc}
\usepackage{amssymb}
\usepackage[margin=1in]{geometry}
\usepackage{amsmath}
\usepackage{graphicx}
\usepackage[table,xcdraw]{xcolor}

\def\trento{T$_\mathrm{R}$ENTo}

\newcommand{\Mev}[0]{M_\textrm{ev}}
\newcommand{\Mevtruth}[0]{\Mev^\textrm{truth}}

\begin{document}

\title{Computational budget optimization for Bayesian parameter estimation in heavy ion collisions}
\date{August 2021}
\author{Brandon Weiss}
\affiliation{Department of Physics, Duke University, Durham NC 27708.}
\author{Jean-Fran\c{c}ois Paquet}
\affiliation{Department of Physics and Astronomy, Vanderbilt University, Nashville TN 37235.}
\affiliation{Department of Mathematics, Vanderbilt University, Nashville TN 37235.}
\affiliation{Department of Physics, Duke University, Durham NC 27708.}
\author{Steffen A. Bass}
\affiliation{Department of Physics, Duke University, Durham NC 27708.}

\begin{abstract}
Bayesian parameter estimation provides a systematic approach to compare heavy ion collision models with measurements, leading to constraints on the properties of nuclear matter with proper accounting of  experimental and theoretical uncertainties.
Aside from statistical and systematic model uncertainties, interpolation uncertainties can also play a role in Bayesian inference, if the model's predictions can only be calculated at a limited set of model parameters.
This uncertainty originates from using an emulator to interpolate the model's prediction across a continuous space of parameters. 
In this work, we study the trade-offs between the emulator (interpolation) and statistical uncertainties.
We perform the analysis using spatial eccentricities from the \trento{} model of initial conditions for nuclear collisions.
Given a fixed computational budget, we study the optimal compromise between the number of parameter samples and the number of collisions simulated per parameter sample. For the observables and parameters used in the present study, we find that the best constraints are achieved when the number of parameter samples is slightly smaller than the number of collisions simulated per parameter sample.
\end{abstract}

\maketitle

\section{Introduction}



One of the main goals of the heavy ion program pursued at the Relativistic Heavy Ion Collider (RHIC) and the Large Hadron Collider (LHC) is a quantitative understanding of the  properties of nuclear matter under extreme conditions as described by Quantum Chromodynamics (QCD). At sufficiently high temperatures and densities accessible during these nuclear collisions, a transient state of matter, the quark-gluon plasma (QGP), is formed, but decays again as the collision systems cools down and disintegrates. 
Collisions of heavy ions leading to QGP formation are multistage processes: the initial impact of the nuclei, the formation and expansion of the QGP, reconfinement into hadrons, and subsequent hadronic interactions~\cite{Bass:2001gb,Gale:2013da,Heinz:2013th,DerradideSouza:2015kpt}.
A wide range of properties of nuclear matter enter numerical simulations of heavy ion collisions. Some of these properties are constrained by external means, for example lattice Quantum Chromodynamics calculations of the equation of state of nuclear matter~\cite{Philipsen:2012nu}.
Other properties, such as transport coefficients or parameters entering the description of the early stage of the collision, are often parametrized and constrained by comparison with data.

Measurements from the Relativistic Heavy Ion Collider (RHIC) and the Large Hadron Collider (LHC) can constrain these physical parameters. Because of the heterogeneity and large number of measurements and model parameters, it is beneficial to perform model-to-data comparisons using statistical techniques such as Bayesian parameter estimation: this allows for a systematic propagation of uncertainties from experimental measurements to physical parameters. A number of such studies have been performed over the past decade~\cite{Petersen:2010zt, Novak:2013bqa, Sangaline:2015isa, Pratt:2015zsa, Bernhard:2015hxa, Bernhard:2016tnd, Moreland:2018gsh, Bernhard:2018hnz, Bernhard:2019bmu, JETSCAPE:2020mzn, Nijs:2020ors, Nijs:2020roc,  JETSCAPE:2020shq,Parkkila:2021tqq,Parkkila:2021yha}.

A key ingredient of Bayesian parameter estimation as used in heavy ion collisions is emulation: given the model's prediction at a \emph{discrete} sample of parameters, an emulator will interpolate the model's prediction over a \emph{continuous} range of parameters. Furthermore, emulators such as Gaussian processes provide an estimate of their own interpolation uncertainty; in effect, this allows emulation to be used in model-to-data comparisons even if the emulator's interpolation uncertainty is not negligible compared to the other uncertainties in the problem --- a situation that is often unavoidable in simulations with large number of parameters. In Bayesian parameter estimation, this emulator uncertainty is included into the physical parameter's final uncertainty determination, alongside experimental, statistical and other theoretical uncertainties.

The most straightforward approach to reducing emulator uncertainty is to increase the number of parameter samples at which the model observables are evaluated; these parameter samples are generally referred to as the emulator's ``design points''. Evidently, this increase in the number of design points must be balanced with other demands on the available computational budget. In particular, simulations of heavy ion collisions are stochastic: there are event-by-event fluctuations in the outcome of the collisions. Comparisons with measurements require simulating and averaging over a large number of collisions. At a fixed computational budget, one must determine the proper trade-off between reducing emulator interpolation uncertainties or statistical ones. In this work, we use the \trento{} model of initial conditions for heavy ion collisions~\cite{Moreland:2014oya} to study this trade-off between statistical and emulator uncertainties.

\section{Methodology}

\label{sec:methodology}

We use the model \trento{} as proxy for observables in heavy ion collisions; this is based on the known correlation~\cite{Luzum:2013yya} between (i) initial spatial anisotropies of models like \trento{}, and (ii) measurable momentum anisotropies of final state hadrons. To interpolate the output of the \trento{} model, we use Gaussian process emulators, which have been the standard emulation technique used for model-to-data comparisons in heavy ion physics~\cite{Petersen:2010zt, Novak:2013bqa, Sangaline:2015isa, Bernhard:2015hxa}. To quantify the optimal trade-off between statistical and emulator uncertainty, we use ``closure tests'', which apply Bayesian parameter inference to model calculations. We summarize the overall approach and methods below.

\subsection{\trento}

\trento{} is a parametric initial condition ansatz for simulating relativistic heavy ion collisions~\cite{Moreland:2014oya}. 
It takes as input the type of colliding nuclei and the inelastic nucleon-nucleon cross-section corresponding to the center-of-mass energy of the collision.

We vary three parameters of the \trento{} model: the effective nucleon size $w$, the fluctuation parameter $k$ and the reduced thickness $p$. The fluctuation parameter $k$ allows for variation in the amount of energy carried by each nucleon, while the reduced thickness $p$ parametrize how the energy or entropy density is constructed from the profile of nucleons sampled from each nuclei.

The output of \trento{} is a density profile in the plane transverse to the collision axis. We interpret \trento{}'s output as an energy density profile $\epsilon(r,\phi)$ with arbitrary normalization. From this density profile $\epsilon(r,\phi)$, we compute single-event spatial anisotropies $\varepsilon_n$~\cite{Luzum:2013yya}:
\begin{equation}
    \varepsilon_n e^{i n \Phi_n} = \frac{\int _0^{\infty } d r r  \int _0^{2 \pi} d\phi \; r^n  \epsilon(r,\phi) e^{i n \phi}}{\int _0^{\infty } d r r  \int _0^{2 \pi} d\phi  \; r^n \epsilon(r,\phi)}
\end{equation}

Although these anisotropies can be related to the momentum anisotropy of hadrons measured at the end of the collision~\cite{Ollitrault:1992bk,Miller:2003kd,PHOBOS:2006dbo,Alver:2010gr,Gardim:2011xv}, in this work, we focus on the initial spatial eccentricies themselves.

We compute the average of an ensemble of \trento{} events with an arithmetic average:
\begin{equation}
    \langle \varepsilon_n \rangle = \frac{1}{\Mev} \sum_{j=1}^{\Mev} \varepsilon_n\{\textrm{event } j\} \; .
    \label{eq:averaged_epsilon_n}
\end{equation}
We used up to four harmonics in this work, $n=2$ to $5$. 

We studied minimum bias Pb-Pb collisions with Woods-Saxon nucleon distributions. We used an inelastic nucleon-nucleon cross-section of $64$~mb, corresponding approximately to $\sqrt{s_{NN}}=2.76$~TeV collisions.


\subsection{Bayesian parameter estimation}

We note  $y_{\textrm{th},j}(\mathbf{p})$ the model's prediction for the ``$j$''-th observables of interest, evaluated when the model's parameters are set to $\mathbf{p}$. At $\mathbf{p}$, we compute $y_{\textrm{th},j}(\mathbf{p})=\langle \varepsilon_{n_j} \rangle (\mathbf{p})$, where the index $j$ runs over all the observables of interest, which in our case are the different ``$n$'' in Eq.~\ref{eq:averaged_epsilon_n}.

In general, given this information, we want to find the probability that parameters $\mathbf{p}$ are consistent with the data $\{ y_{\textrm{exp},j} \}$ and with the experimental and theoretical uncertainties $\{ \sigma_{\textrm{exp},j} \}$ and $\{ \sigma_{\textrm{th},j}(\mathbf{p}) \}$. The first step is to define a metric to quantify the level of model-to-data agreement: the likelihood function. We make the typical choice of a Gaussian likelihood function:
\begin{equation}
    \mathcal{L}(\{ y_{\textrm{exp},j} \} |\mathbf{p}) = \left.
    \exp\left(-\frac{1}{2} \sum_j \frac{\left[ y_{\textrm{th},j}(\mathbf{p}) - y_{\textrm{exp},j} \right]^2} {\sigma_{\textrm{th},j}(\mathbf{p})^2 + \sigma_{\textrm{exp},j}^2}\right) \middle/ \sqrt{ (2\pi)^n \prod_j \left (\sigma_{\textrm{th},j}(\mathbf{p})^2 + \sigma_{\textrm{exp},j}^2\right) } \right. \; .
\end{equation}
We assumed a diagonal covariance matrix. In this work, instead of comparing the model with data, we will use closure tests: we will replace $y_{\textrm{exp},j}$ by model calculations, as discussed later. These model calculations do stand for experimental data, and consequently we keep the label ``exp'' to denote these quantities.

Given our model of the collision, the probability that the parameter set $\mathbf{p}$ is consistent with the data and the uncertainties is given by the posterior distribution of model parameters, $\mathcal{P}(\mathbf{p} | (\{ y_{\textrm{exp}} \})$:
\begin{equation}
\mathcal{P}(\mathbf{p} | (\{ y_{\textrm{exp}} \}) = \frac{\mathcal{L}(\{ y_{\textrm{exp}} \} |\mathbf{p}) \textrm{Prior}(\mathbf{p})}{\int d\mathbf{p} \mathcal{L}(\{ y_{\textrm{exp}} \} |\mathbf{p}) \textrm{Prior}(\mathbf{p}) }
\label{eq:posterior}
\end{equation}
where $\textrm{Prior}(\mathbf{p})$ is the prior distribution. In this work we take the prior to be constant over a finite parameter range, for all the parameters.

The posterior distribution $\mathcal{P}(\mathbf{p} | (\{ y_{\textrm{exp}} \})$ is a probability distribution whose dimensionality is equal to the number of model parameters. Different projections of the posterior distribution summarize the constraints on the model parameters. As long as the number of parameters is small and the model is relatively fast computationally, it is straightforward to sample and marginalize the posterior $\mathcal{P}(\mathbf{p} | (\{ y_{\textrm{exp}} \})$. When the number of parameters increases or if the model is expensive, it can become overly burdensome to evaluate a model's prediction at a large number of values of the parameter $\mathbf{p}$ to compute $\mathcal{P}(\mathbf{p} | (\{ y_{\textrm{exp}} \})$. A solution is to prepare a fast proxy, such as a Gaussian process~\cite{williams2006gaussian}, to emulate the model's prediction. We review Gaussian process emulators briefly in the next subsection. As far as the Bayesian inference is concerned, the consequence of using Gaussian process emulators is substituting the model's prediction by the emulator's prediction,
\begin{equation}
y_{\textrm{th},j}(\mathbf{p})  \to y_{\textrm{emul},j}(\mathbf{p}) \; ,
\end{equation}
as well as adding the emulator interpolation uncertainty into the sum of uncertainties,
\begin{equation}
    \sigma_{\textrm{th},j}(\mathbf{p})^2 \to \sigma_{\textrm{th},j}(\mathbf{p})^2 + \sigma_{\textrm{emul},j}(\mathbf{p})^2 \; .
\end{equation}






\subsection{Emulation with Gaussian processes}

\begin{figure}[tb]
    \centering
    \includegraphics[scale=0.9]{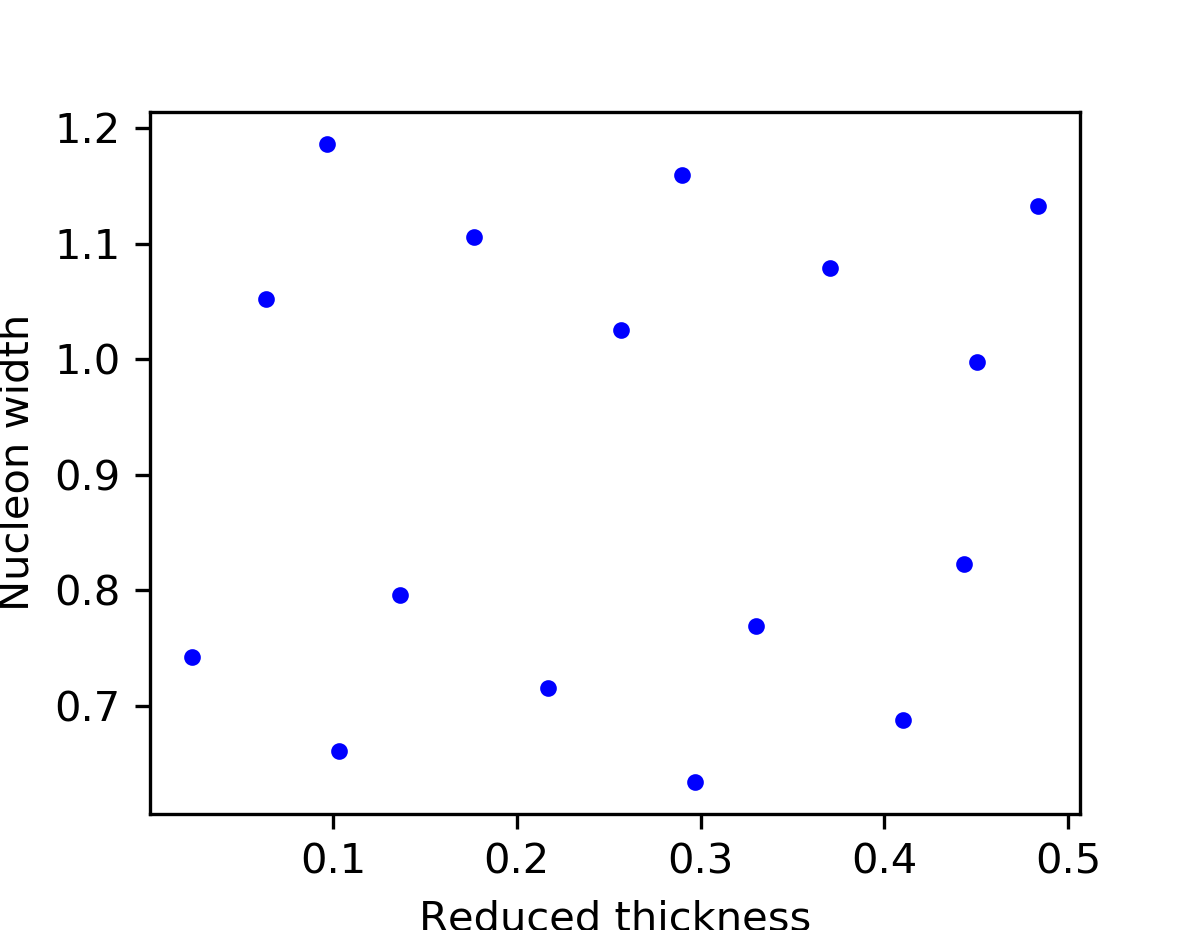}
    \caption{16 design points in a two-dimensional parameter space (reduced thickness on the x-axis and nucleon width on the y-axis), as sampled with  a low discrepancy sequence generator~\cite{low_discrepancy,niederreiter1992random}.}
    \label{fig:lattice}
\end{figure}

The idea behind emulation is to first sample the parameter space of the model, $\{ \mathbf{p}_k \}$; the parameter samples are the design points. The model's predictions are then calculated for all design points: $\{ y_{\textrm{th},j}(\mathbf{p}_k) \}$. The emulator then interpolates the model's predictions over a continuous range of model parameters $\mathbf{p}$.

To sample the parameter space, we use a low discrepancy sequence generator~\cite{low_discrepancy,niederreiter1992random}. Figure~\ref{fig:lattice} shows an example for 16 design points within a two-dimensional parameter space.

\begin{figure}[tb]
    \centering
    \includegraphics[scale=0.5]{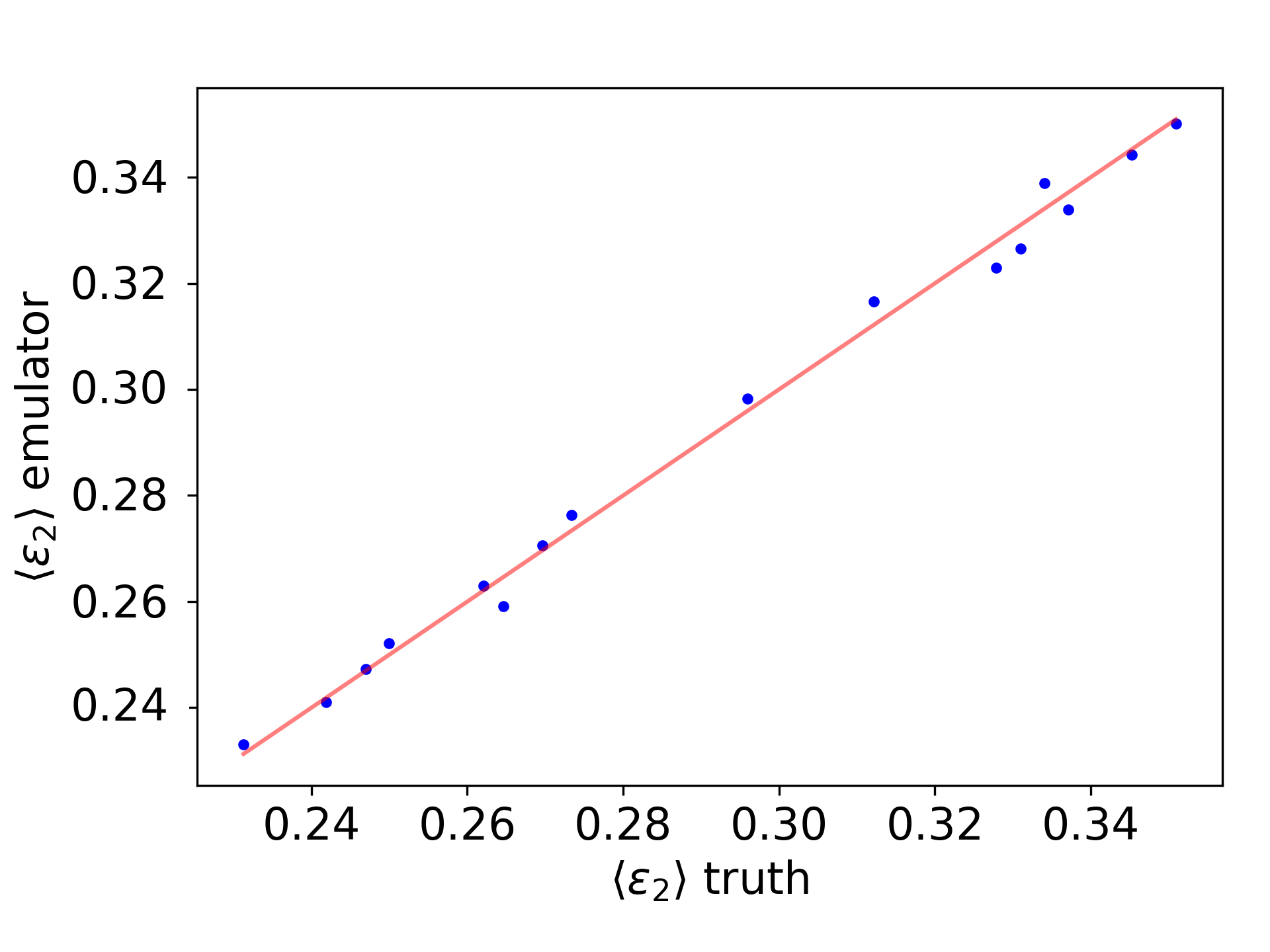}
    \includegraphics[scale=0.5]{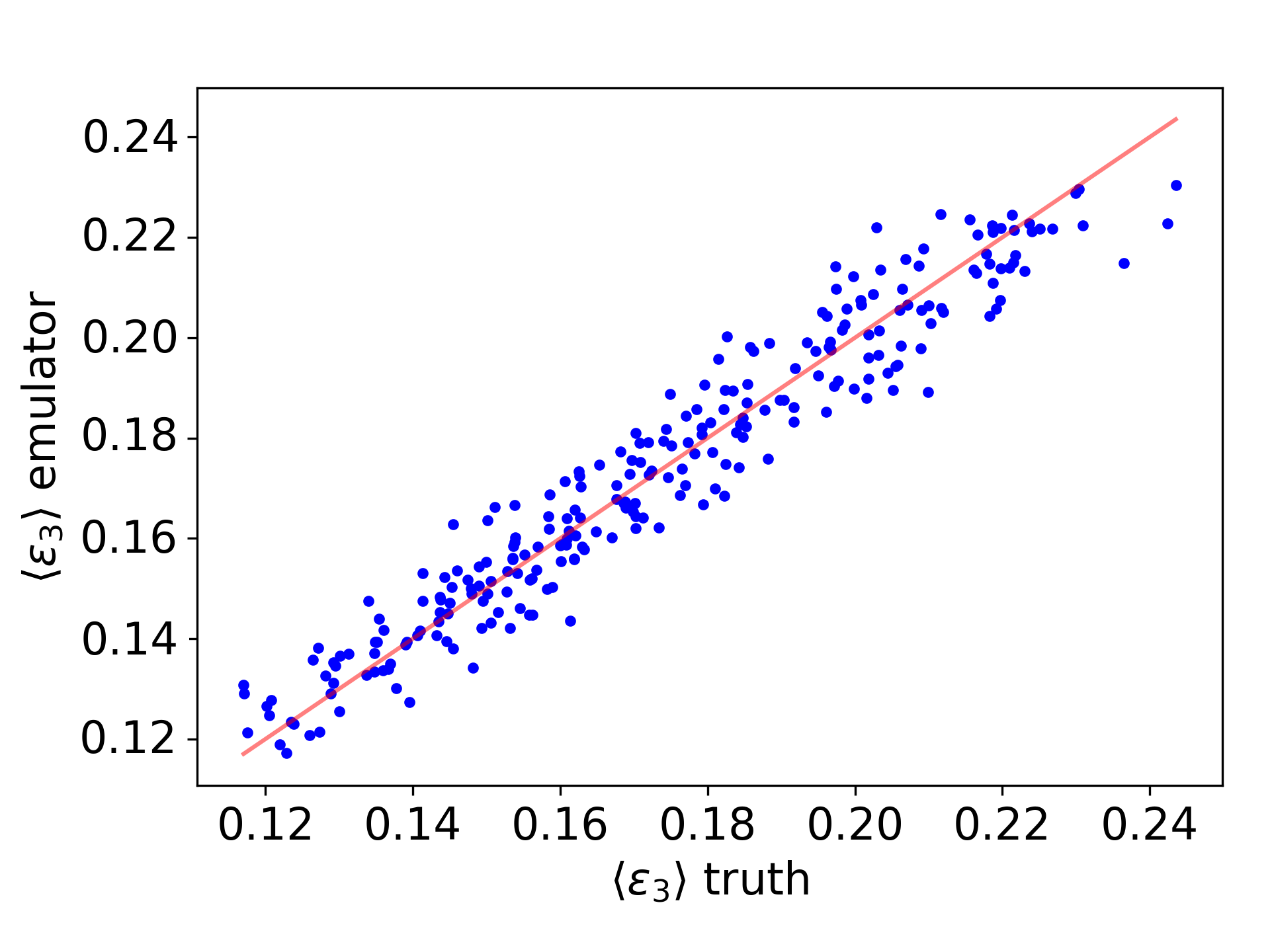}
    \caption{(Left) Value for $\langle \varepsilon_{2} \rangle$ as predicted by a Gaussian process emulator trained with 16 design points (y-axis), compared to the actual value of the model at the design points (x-axis); (Right) Same as left, for $\langle \varepsilon_{3} \rangle$ with 256 design points. The red line corresponding to perfect emulation, $y_{\textrm{emul},j}(\mathbf{p})=y_{\textrm{th},j}(\mathbf{p})$, is shown for reference. Statistical uncertainties on each sample are not shown for clarity.}
    \label{fig:closure1}
\end{figure}

For each observable ($\langle \epsilon_n \rangle$), a Gaussian process is trained on the set of values of $\langle \epsilon_n \rangle(\mathbf{p})$ calculated at the design points $\{ \mathbf{p}_k \}$. Gaussian process emulators are probabilistic interpolators, which assume that each point in parameter space is a probability distribution. At the design points, the width of the probability distribution should include the statistical uncertainty of the observables. Between the design points, the width of the probability distribution should increase to account for the interpolation uncertainty.
To handle stochastic simulations, Gaussian process emulators must be set up such that variations between parameter samples are divided into  (i) statistical uncertainty, and (ii) the actual variation from the parameter dependence of the model. A mathematical description of the approach can be found in Ref.~\cite[Section~V-B-2]{JETSCAPE:2020mzn}. In short, a white noise kernel accounts for the uncorrelated (``short range'') fluctuations originating from the statistical uncertainty, and a squared-exponential kernel accounts for the longer-range parameter dependence. The parameters of these kernels, for example the relative size of the kernels, is determined by numerical optimization. This approach is used in most applications of Bayesian parameter inference in heavy ion physics. The result of this implementation is that the emulator uncertainty accounts for both the statistical and interpolation uncertainties.

Figure~\ref{fig:closure1} shows the mean-value predictions of the emulator for $\langle \epsilon_2 \rangle$ and $\langle \epsilon_3 \rangle$ compared to the actual value of these observables. The points all lie close to the line $y_{\textrm{emul},j}(\mathbf{p})=y_{\textrm{th},j}(\mathbf{p})$, which shows that the emulator does indeed interpolate well. On the other hand, $\langle \epsilon_3 \rangle$ tends to have a larger statistical uncertainty than $\langle \epsilon_2 \rangle$, and as the right-hand side of Figure~\ref{fig:closure1} shows, increasing the number of design points will not lead to perfect deterministic-like prediction from the emulator because the values of $\langle \epsilon_3 \rangle$ have significant statistical uncertainties. This is the trade-off that we focus on in this work.


\subsection{Closure tests}

\label{sec:methodology_closure}

Closure tests are a straightforward application of Bayesian parameter inference, with experimental data replaced by calculations from the model itself. It is a self-consistency check made non-trivial by (i) the presence of uncertainties, as well as (ii) loss of information from the model to the observables. A longer discussion of closure tests, in the context of heavy ion collisions, can be found in Section~VI of Ref.~\cite{JETSCAPE:2020mzn}. We describe it briefly here, with focus on metrics to quantify the success of closure tests.

In closure tests, we first select a set of parameters that are considered the reference parameter for the closure test; we refer to these parameters as the ``reference'' or ``truth'' parameters, $\mathbf{p}^{\textrm{truth}}$. We then compute event-averaged eccentricities $\langle \epsilon^{\textrm{truth}}_n \rangle=\langle \varepsilon_n \rangle(\mathbf{p}^{\textrm{truth}})$, which have a statistical uncertainty $\Delta \langle \epsilon^{\textrm{truth}}_n \rangle$.  While these uncertainties are purely statistical, they are meant to mimic statistical and systematic uncertainties found in measurements. This uncertainty can be dialed by changing the number of \trento{} events $\Mevtruth$ used to compute $\langle \epsilon^{\textrm{truth}}_n \rangle$.

Separately, we train Gaussian process emulators for $\langle \varepsilon_n \rangle(\mathbf{p})$ over a chosen range of model parameters, as described in the previous section. The two sources of uncertainties in the emulator are the number of design points (parameter samples)  $N_d$, which controls the interpolation uncertainty, and the number of \trento{} events per design point, $\Mev$, which control the statistical uncertainty of $\langle \varepsilon_n \rangle(\mathbf{p})$ at each design point.

Using the Gaussian process emulators as proxy for \trento{}, we then perform Bayesian parameter estimation to attempt to recover the parameters $\mathbf{p}^{\textrm{truth}}$ from the values of the observables $\langle \epsilon^{\textrm{truth}}_n \rangle \pm \Delta  \langle \epsilon^{\textrm{truth}}_n \rangle$. The result of the Bayesian inference is the posterior $\mathcal{P}(\mathbf{p} | \{\langle \epsilon^{\textrm{truth}}_n \rangle\})$ (Eq.~\ref{eq:posterior}). This posterior distribution depends on three sources of uncertainties: the uncertainty on the reference observables ($\Delta \langle \epsilon^{\textrm{truth}}_n \rangle$), which mimics experimental uncertainty, and the statistical and interpolation uncertainty in the emulator.
If all uncertainties were small, one would expect the posterior to be a narrow peak at the parameter truth $\mathbf{p}^{\textrm{truth}}$. In practice, some of these uncertainties are significant, and the posterior distribution thus has some finite width in parameter space. If closure is successful, the posterior should enclose the parameter set $\mathbf{p}^{\textrm{truth}}$. 


\begin{figure}[tb]
    \centering
    \includegraphics[scale=0.6]{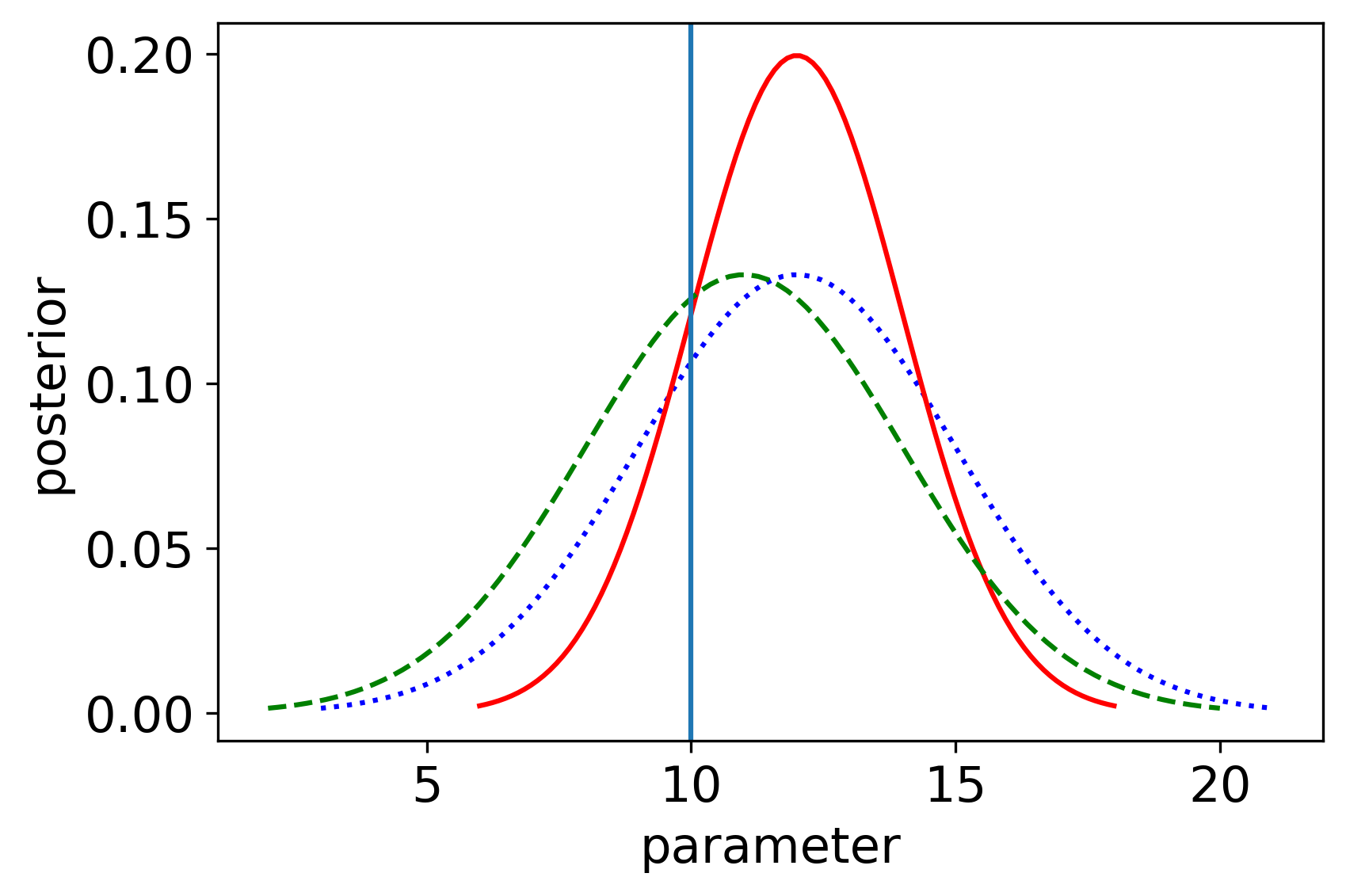}
    \caption{Schematic comparison of 1-parameter posterior distributions (red solid, green dashed and blue dotted Gaussians) with a single fixed ``truth'' value of the parameter (vertical teal line). The green dashed line shows how the posterior increases as \emph{accuracy} increases when compared to the blue dotted line, and the red solid line shows how the posterior increases as \emph{precision} increases when compared to the blue dotted line.  }
    \label{fig:normPost}
\end{figure}

To quantify the degree of agreement of the posterior --- $\mathcal{P}(\mathbf{p} | \{\langle \epsilon^{\textrm{truth}}_n \rangle\})$ --- with the known true value $\mathbf{p}^{\textrm{truth}}$ of the parameters used in the closure test, we use two different metrics. The first one is the value of the posterior at the parameter truth, $\mathcal{P}(\mathbf{p}^{\textrm{truth}} | \{\langle \epsilon^{\textrm{truth}}_n \rangle\})$. The posterior at the parameter truth balances accuracy and precision. As illustrated in Fig.~\ref{fig:normPost}, an increase in accuracy (shift towards true value of the parameter) will increase the posterior value. An increase in precision (narrower posterior) will also increase the posterior at the parameter truth, but only to a certain extent: the posterior at the truth will actually start to decrease if the results get too confident about the incorrect value. The posterior at the truth has the benefit of being simple to interpret and inexpensive to compute.

For completeness, we studied a second metric, the Akaike Information Criterion (AIC)~\cite{akaike1974new}, given by:
\begin{equation}
    \textrm{AIC} \equiv -2\ln(\mathcal{L}_\mathrm{max}) + 2k
\end{equation}
where $\mathcal{L}_\mathrm{max}$ is the maximum likelihood in the parameter space and $k$ is the number of parameters. The Akaike Information Criterion is a metric used for model selection, to help determine how successful a model is at describing measurements given its number of parameters~\cite{Liddle:2007fy,liddle2004many}.
In the case of closure tests, while we know that the emulator and the reference (truth) calculations originate from the same model, the uncertainties in the problem can evidently obscure this fact. We will see in this work the AIC's success at quantifying closure. Note that the AIC is more expensive to compute, as it requires identifying the maximum of the likelihood.


\section{Results}

\label{sec:results}

As discussed in Section~\ref{sec:methodology}, the result of Bayesian parameter inference generally depends on the experimental uncertainty as well as on the statistical and interpolation uncertainty of the emulator. Additional theoretical uncertainties can also play a role~\cite{JETSCAPE:2020mzn}.

Because we use closure tests in this manuscript, the experimental uncertainty is replaced by the statistical uncertainty of the reference (``truth'') calculations, which is controlled by changing the number $\Mevtruth$ of \trento{} events that are averaged over to compute the reference observables. As for the emulator, its uncertainty is determined by the number of samples of the model parameters $N_d$ and the number of collisions simulated per parameter sample $\Mev$.

Tests are performed at a fixed computation budget: the product $N_{\textrm{tot}} = N_d \Mev$ is kept fixed. The set of $(N_d, \Mev)$ pairs is chosen to vary by a factor of 2 (e.g. \{8, 128\} $\rightarrow$ \{16, 64\} $\rightarrow$ \{32, 32\}). We use nine different sets of  $(N_d, \Mev)$, resulting in nine different groups of emulators, each group having one Gaussian process emulator per observable.


As a base case, we first used $N_{\textrm{tot}} = N_d \Mev=2^{16}$ total events for a Pb-Pb collision. 
We used $\Mevtruth=2^{16}$ collisions to compute the reference ``truth'' observables. We begin with only two \trento{} parameters at the time --- nucleon width and reduced thickness in one case, and nucleon width and fluctuation in the other. To test the robustness of the methodology, we later vary the data uncertainty, and the number of parameters and of observables.

\begin{figure}[tb]
    \centering
    \includegraphics[width=0.32\textwidth]{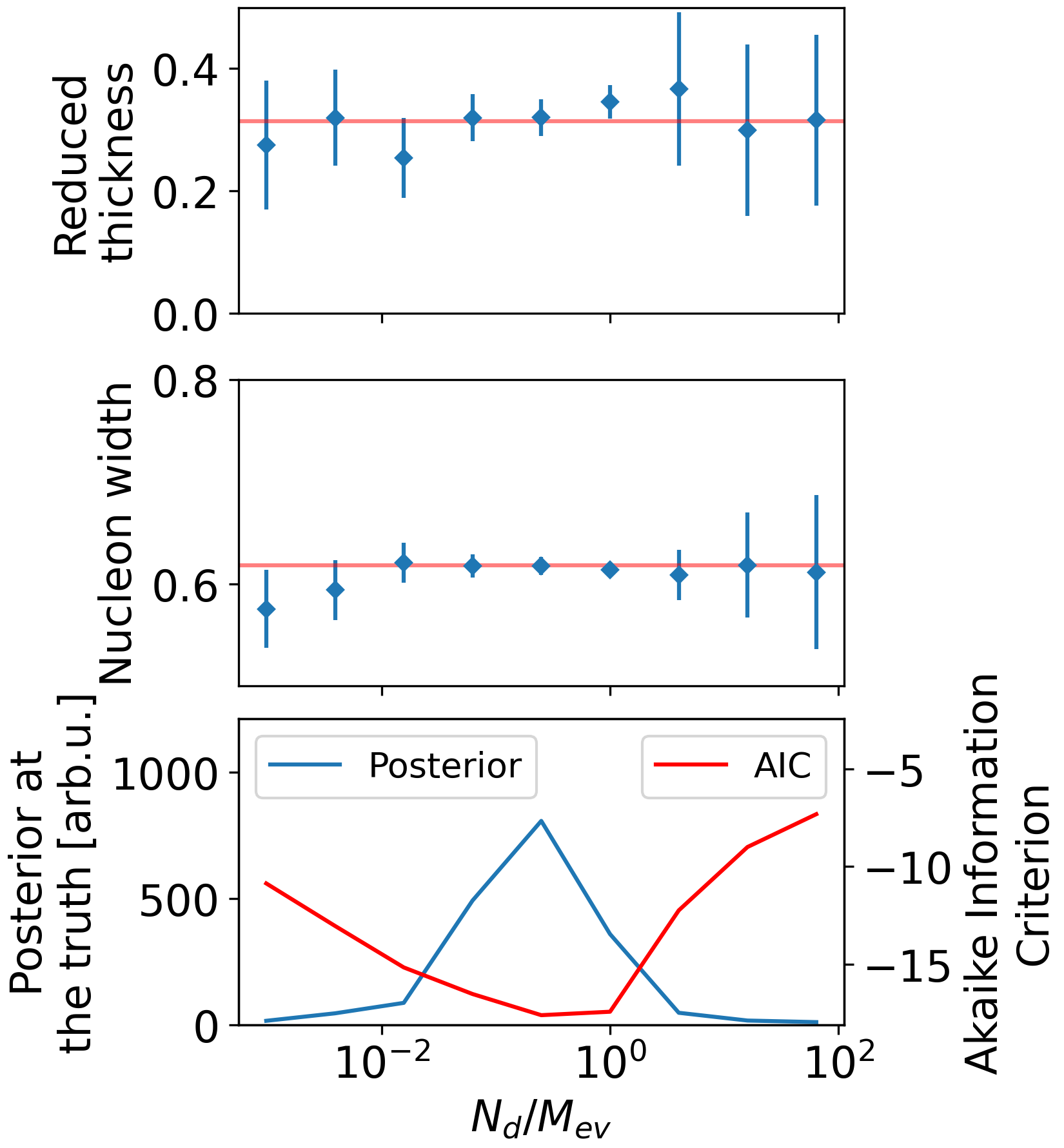}
    \includegraphics[width=0.32\textwidth]{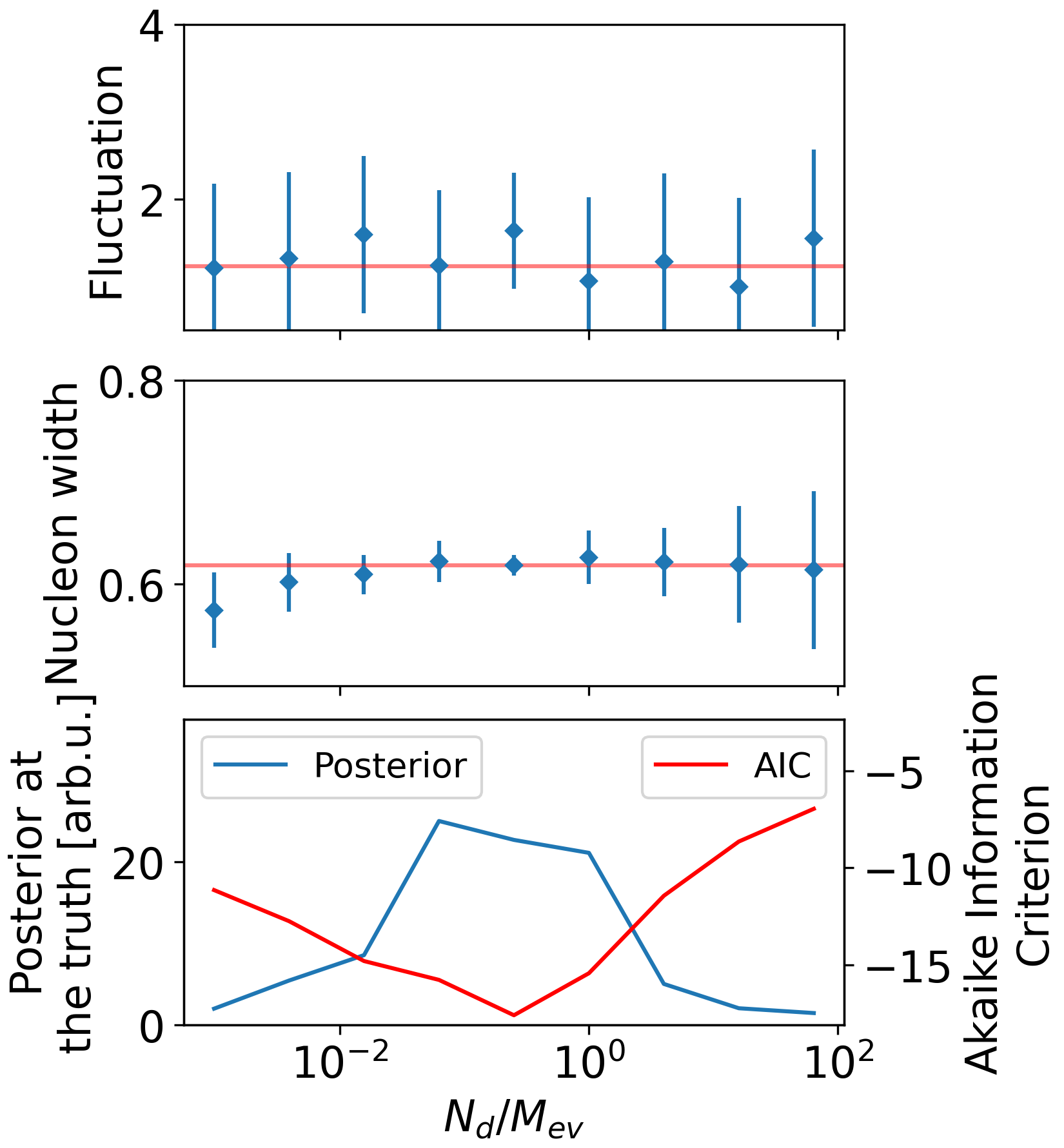}
    \caption{Closure tests with $N_d \Mev = 2^{16}$ total events, $\Mevtruth=2^{16}$ events to compute the reference observables, two parameters (nucleon width and reduced thickness on the left, and nucleon width and reduced fluctuation parameter on the right), and two observables ($\langle \varepsilon_2 \rangle$ and $\langle \varepsilon_3 \rangle$ in Pb-Pb collisions at $\sqrt{s_{NN}}=2.76$~TeV). Top two panels show the mode and width of the one-parameter marginalized likelihood, while the bottom panel shows the value of the posterior at the true value of the parameter as well as the Akaike Information Criterion (AIC), all as a function of the ratio of the number of design points $N_d$ to the number of \trento{} simulations per design points $\Mev$.}
    \label{fig:2to16}
\end{figure}

After performing Bayesian parameter inference in the framework described above, we obtain a 2-parameter posterior distribution. By marginalizing over each parameter, we obtain 1-parameter marginalized posterior distributions, whose mode and width (variance) we calculate and plot as a function of the ratio of the number of design points $N_d$ to the number of \trento{} simulations per design points $\Mev$; this is shown in the left top two panels of Figure~\ref{fig:2to16}. The true (reference) value of the parameters used in the tests is shown as a red horizontal line. We first see that, for any number of design points, the constraints from the posterior are consistent with the true value of the parameters. This is not a trivial observation: it indicates that the emulator is quantifying properly the interpolation and statistical uncertainties, however large or small these uncertainties are. In this sense, we see that Bayesian parameter inference provides valid constraints for any set of design points $N_d$ and number of \trento{} events $\Mev$. On the other hand, one can optimize the results of the Bayesian inference by a careful compromise between emulator and statistical uncertainty. With large numbers of design points (large $N_d/\Mev$), the emulator uncertainty is expected to be small, but large statistical uncertainties on the $\langle \epsilon_n \rangle$ of each design point prevent accurate constraints on the parameter. Using a large number of \trento{} events with a small number of design points (small $N_d/\Mev$) leads to similar sub-optimal results. To quantify this tension, we use the two metrics discussed in Section~\ref{sec:methodology_closure}: the value of the posterior at the true value of the closure test parameters, and the Akaike Information Criterion (AIC). Both are plotted on the bottom left panel of Figure~\ref{fig:2to16}, still as a function of the ratio $N_d/\Mev$. The maximum value of the posterior at the truth, and the minimum value of the AIC, are observed around 
$N_d/\Mev=0.25$, which corresponds to $N_d=2^7$ design points.
Overall, the different panels all indicate that certain apportionment of the number of design points does result in significantly better constraints on the model parameter \emph{for the same computational expense}. The optimal number is found to be between 
$N_d/\Mev=0.1$ and $1$.
This suggests that the optimal value of  $N_d$ is of the order of $\sqrt{N_{\textrm{tot}}}$ or slightly smaller.

Repeating this exercise by swapping one of the \trento{} parameters (reduced thickness) by another (fluctuation parameter) yields the right panels of Figure~\ref{fig:2to16}. Because the fluctuation parameter is more challenging to constrain from $\langle \varepsilon_2 \rangle$ and $\langle \varepsilon_3 \rangle$, the dependence on the number of design point $N_d$ of the top panel is modest. Yet this does not prevent the other parameter to be better constrained (middle right panel), and the optimal number of design points remain in the vicinity of 
$N_d/\Mev=0.25$.


To ensure the robustness of our conclusions, we repeat this same analysis varying the uncertainty on the reference calculations (which are proxy for data uncertainty) as well as varying the number of parameters and observables. 

\subsection{Dependence on the uncertainty of the reference calculation (``data'')}

\begin{figure}[tb]
    \centering
    \includegraphics[width=0.32\textwidth]{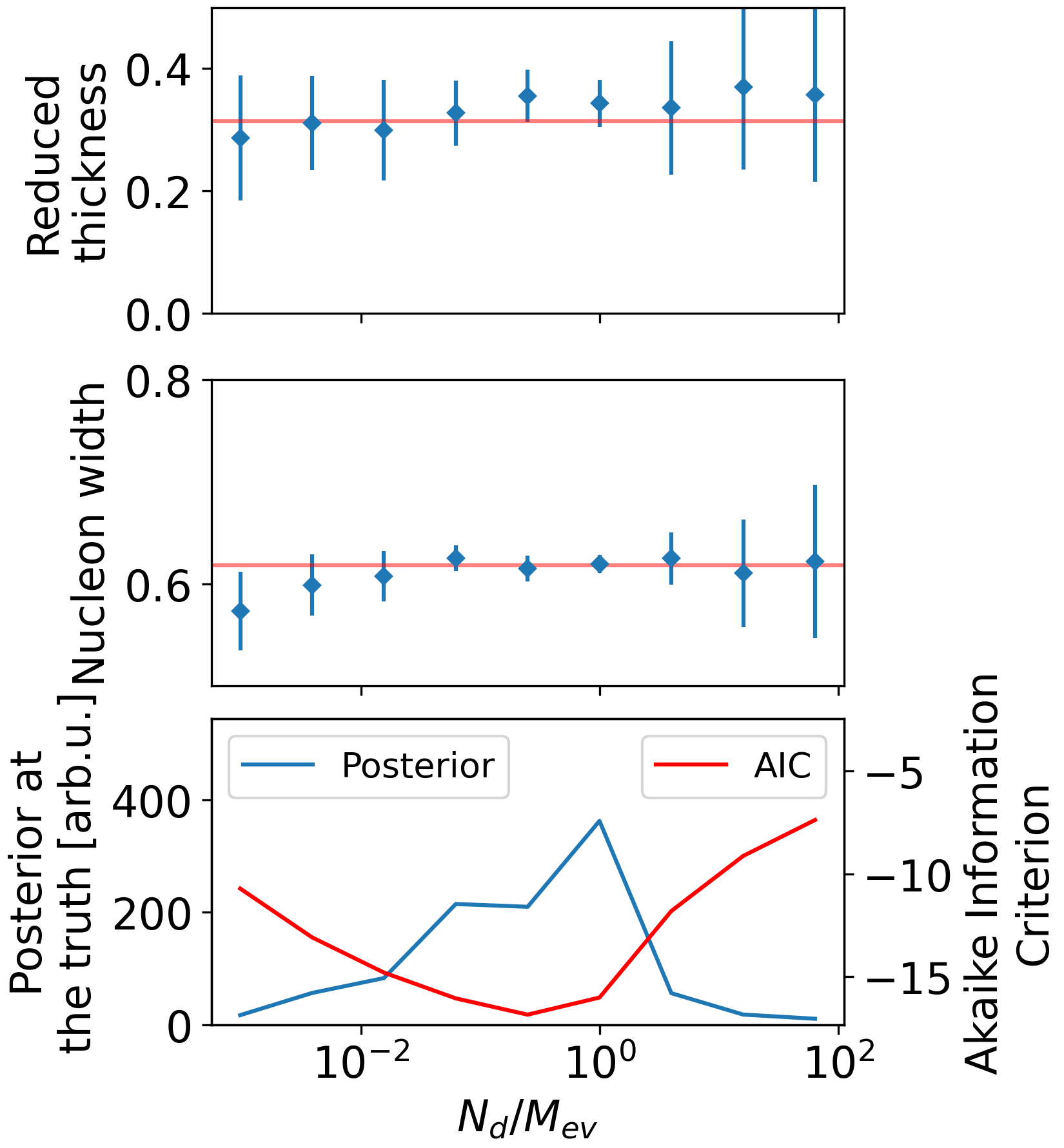}
    \includegraphics[width=0.32\textwidth]{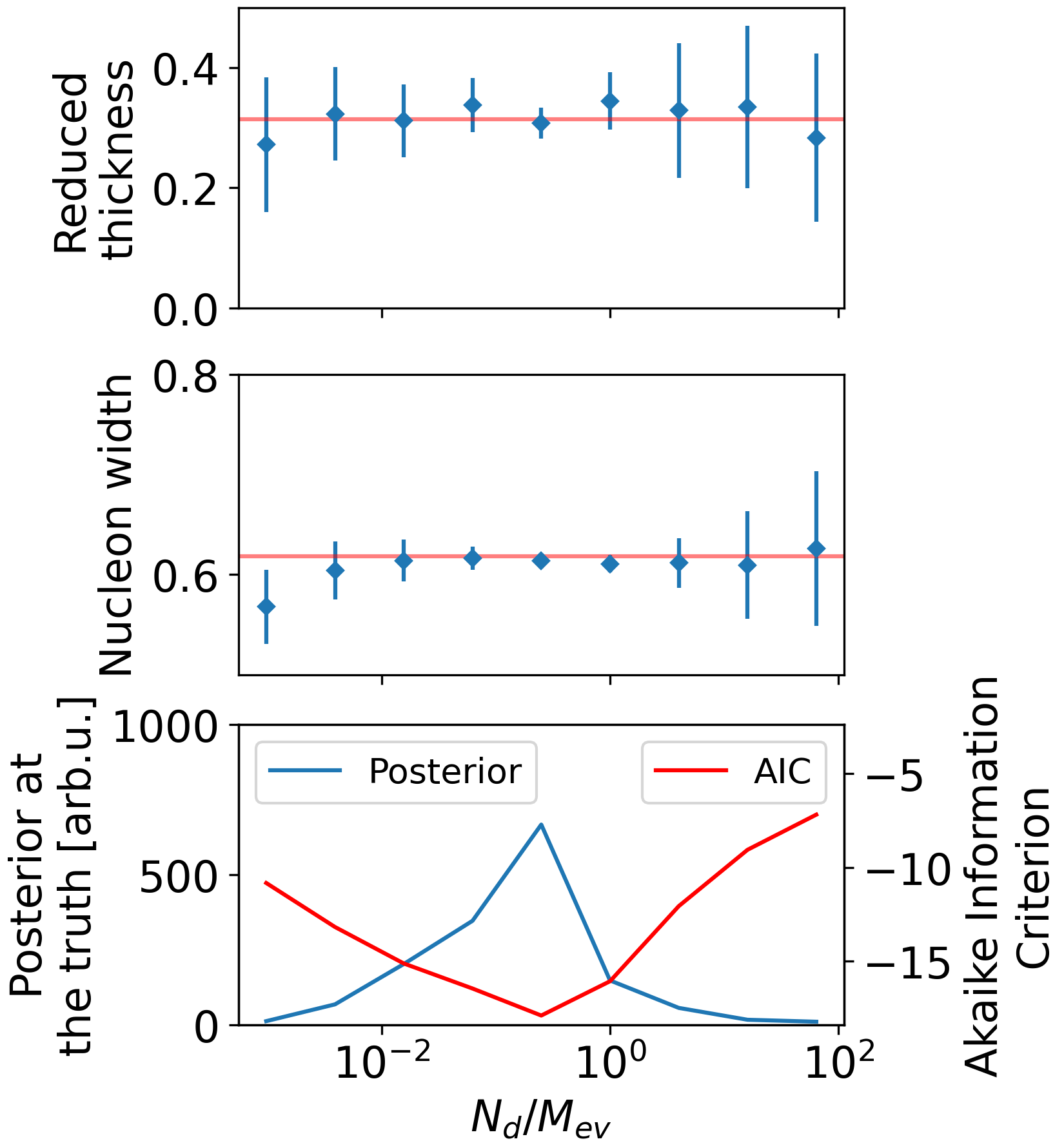}
    \caption{Same as left side of Fig.~\ref{fig:2to16}, with (left) $\Mevtruth=2^{14}$ events and (right) $\Mevtruth=2^{18}$ events.}
    \label{fig:datUncert}
\end{figure}

To compute the reference (``truth'') observables, $\Mevtruth$ \trento{} events are averaged. Changing $\Mevtruth$ controls the uncertainty on the reference observables, $\Delta \langle \epsilon^{\textrm{truth}}_n \rangle$. In this section, we vary $\Mevtruth$ to verify the effect on the optimal number of design points.

In the previous example, because $N_{\textrm{tot}} = N_d \Mev$ and $\Mevtruth$ were both equal to $2^{16}$, the interpolation and statistical uncertainty in the emulator (controlled by $N_{\textrm{tot}}$) were always large compared to the uncertainty of the reference calculations that are stand-in for measurements (controlled by $\Mevtruth$): at best, the statistical uncertainties for the calculations used in the emulator could be equal to the statistical uncertainty from the reference calculations (uncertainty $\propto 1/\sqrt{2^{16}}$) when $N_d=1$, which would evidently be a case with extreme interpolation uncertainty.

We first repeat the previous example with $\Mevtruth=2^{14}$ events while keeping $N_{\textrm{tot}} = N_d \Mev=2^{16}$. This means that, when $N_d=2^2=4$, the statistical uncertainty of the reference calculation is similar to that in the calculations used for the emulator; when $N_d=2^4=16$, the statistical uncertainty in the emulator is approximately a factor $2$ larger than in that of the reference calculations, etc. The results of the closure tests are shown on the left panel of Fig.~\ref{fig:datUncert}. Overall, the results are similar: the constraints on the model parameters are consistent with the true value of the parameters at any ratio $N_d/\Mev$, but the range $N_d/\Mev\approx 0.1$--$1$ is found to be optimal.

We repeat this test in the opposite direction, using $\Mevtruth=2^{18}$ events to reduce the uncertainty on the reference calculations while keeping the emulator budget the same ($N_{\textrm{tot}} = N_d \Mev=2^{16}$). This is  equivalent to a scenario where the emulator uncertainties are always being much larger than the data uncertainties. This yields the right panel of Fig.~\ref{fig:datUncert}. Again, the optimal range of design points remains $N_d/\Mev\approx 0.1$--$1$.




\subsection{Dependence on the observables and the parameters}

\begin{figure}[tb]
    \centering
    \includegraphics[scale=0.49]{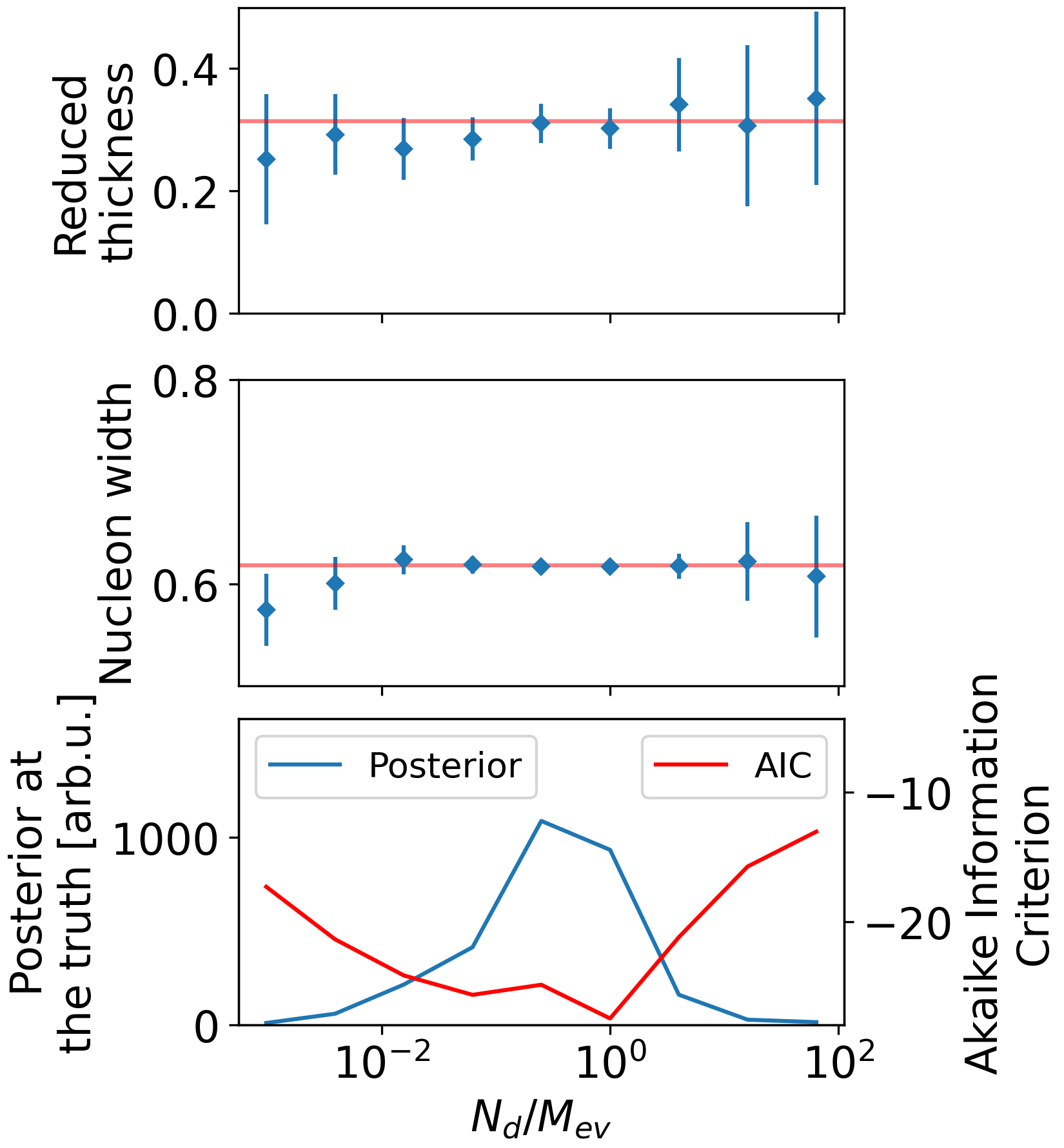}
    \includegraphics[scale=0.49]{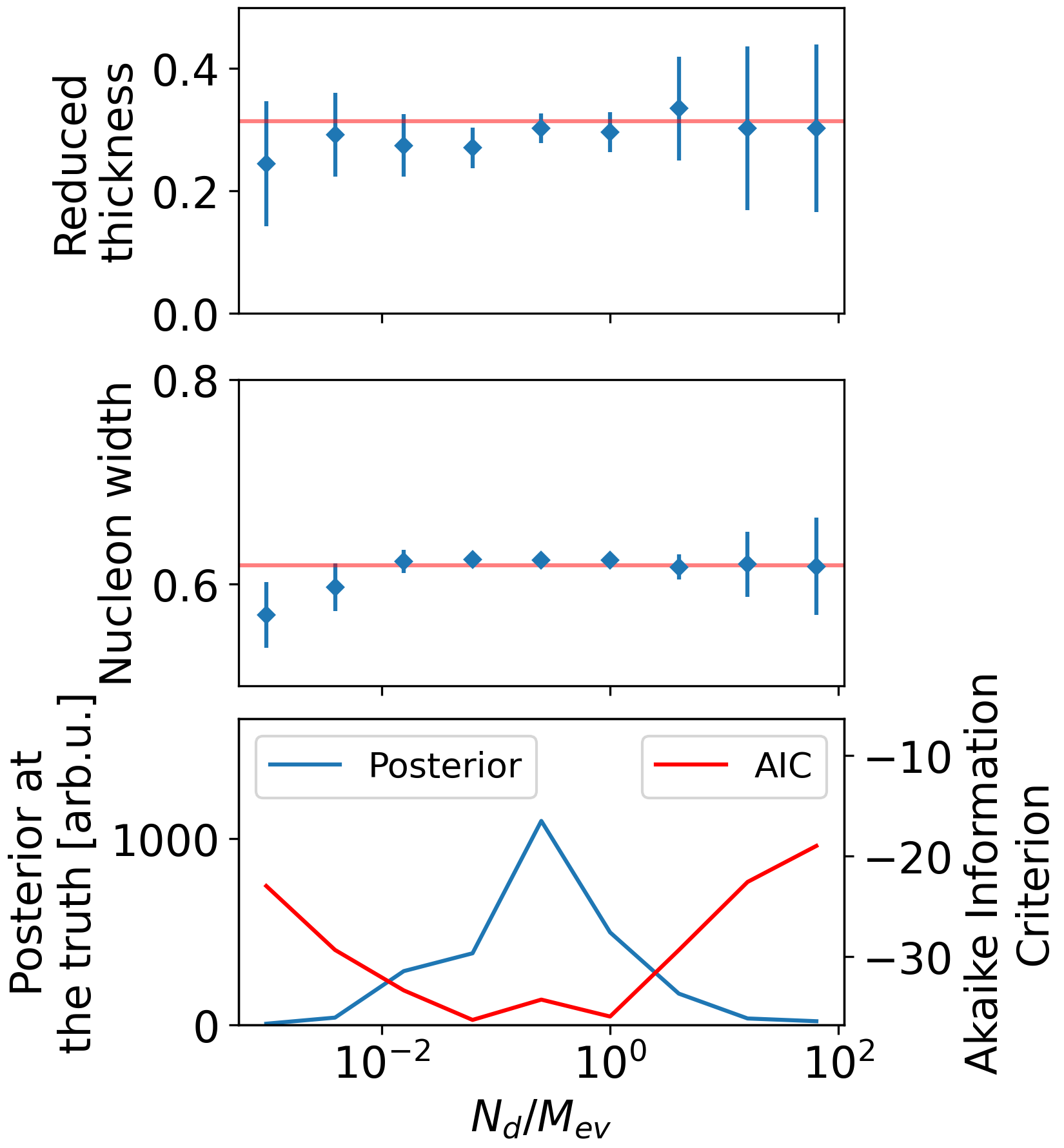}
    \caption{Same as left side of Fig.~\ref{fig:2to16}, with (left) three observables ($\langle \varepsilon_n \rangle$, $n=2,3,4$), and (right) four observables ($\langle \varepsilon_n \rangle$, $n=2,3,4,5$).}
    \label{fig:e4}
\end{figure}

\begin{figure}[tbp]
    \centering
    \includegraphics[width=0.32\textwidth]{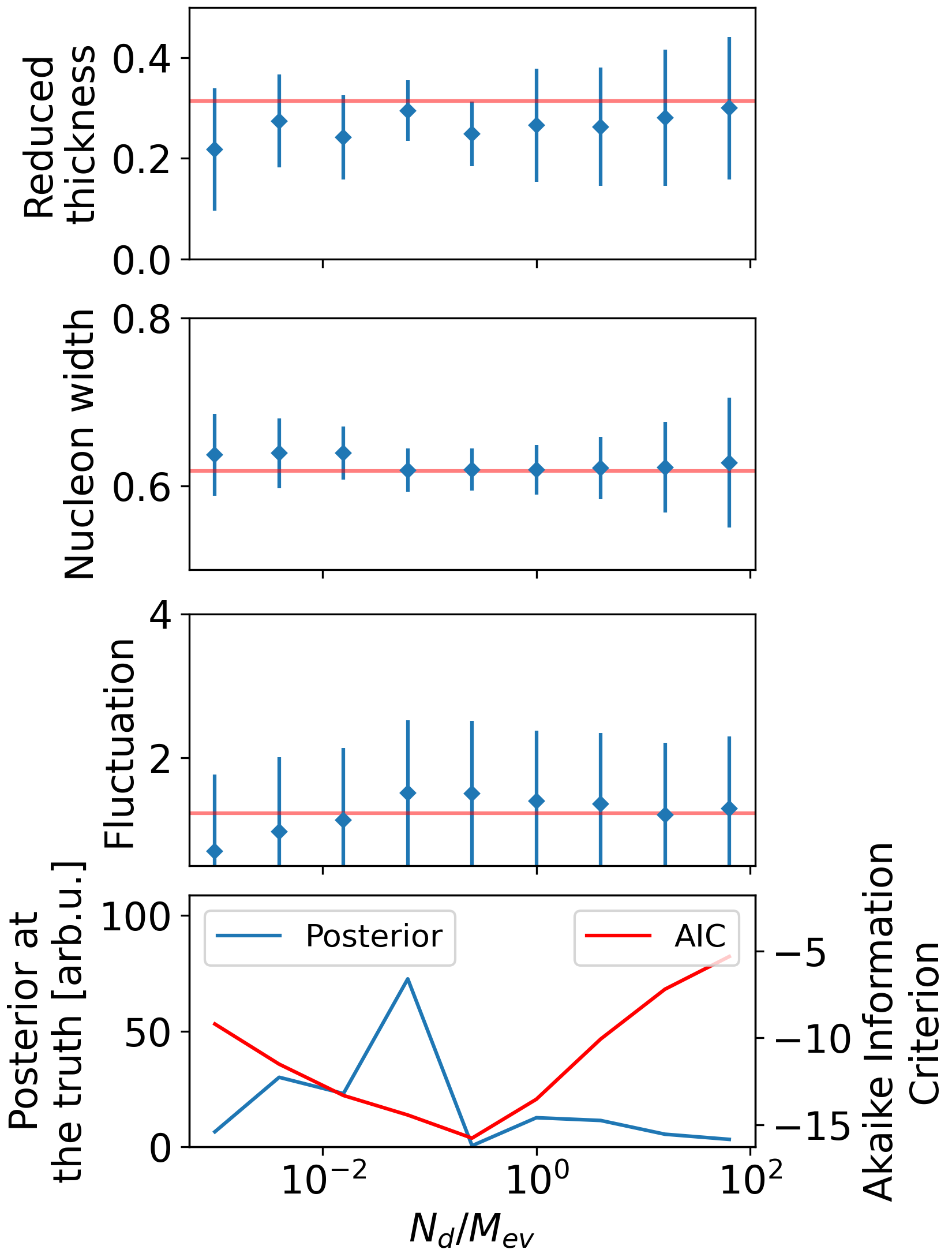}
    \includegraphics[width=0.32\textwidth]{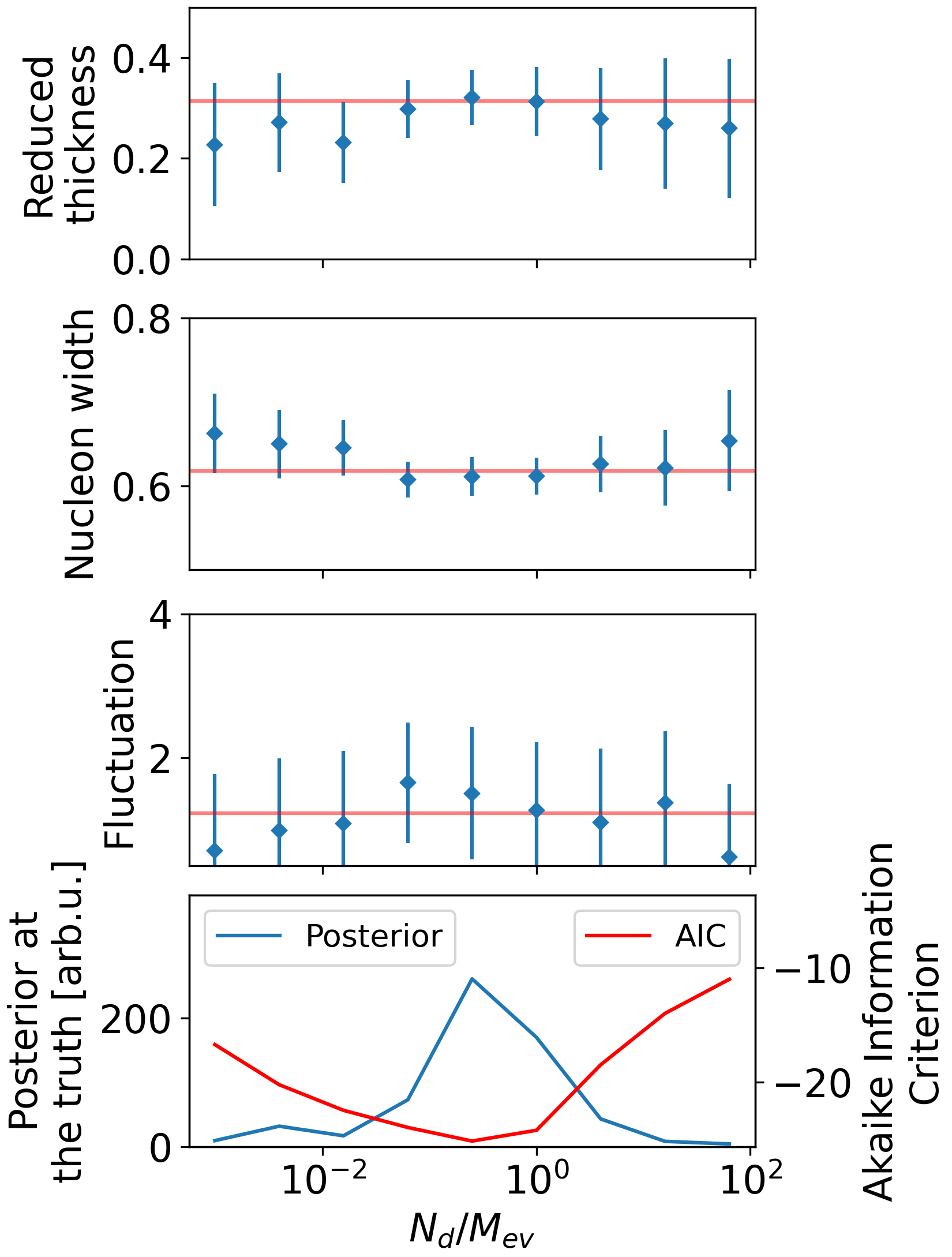}
    \includegraphics[width=0.32\textwidth]{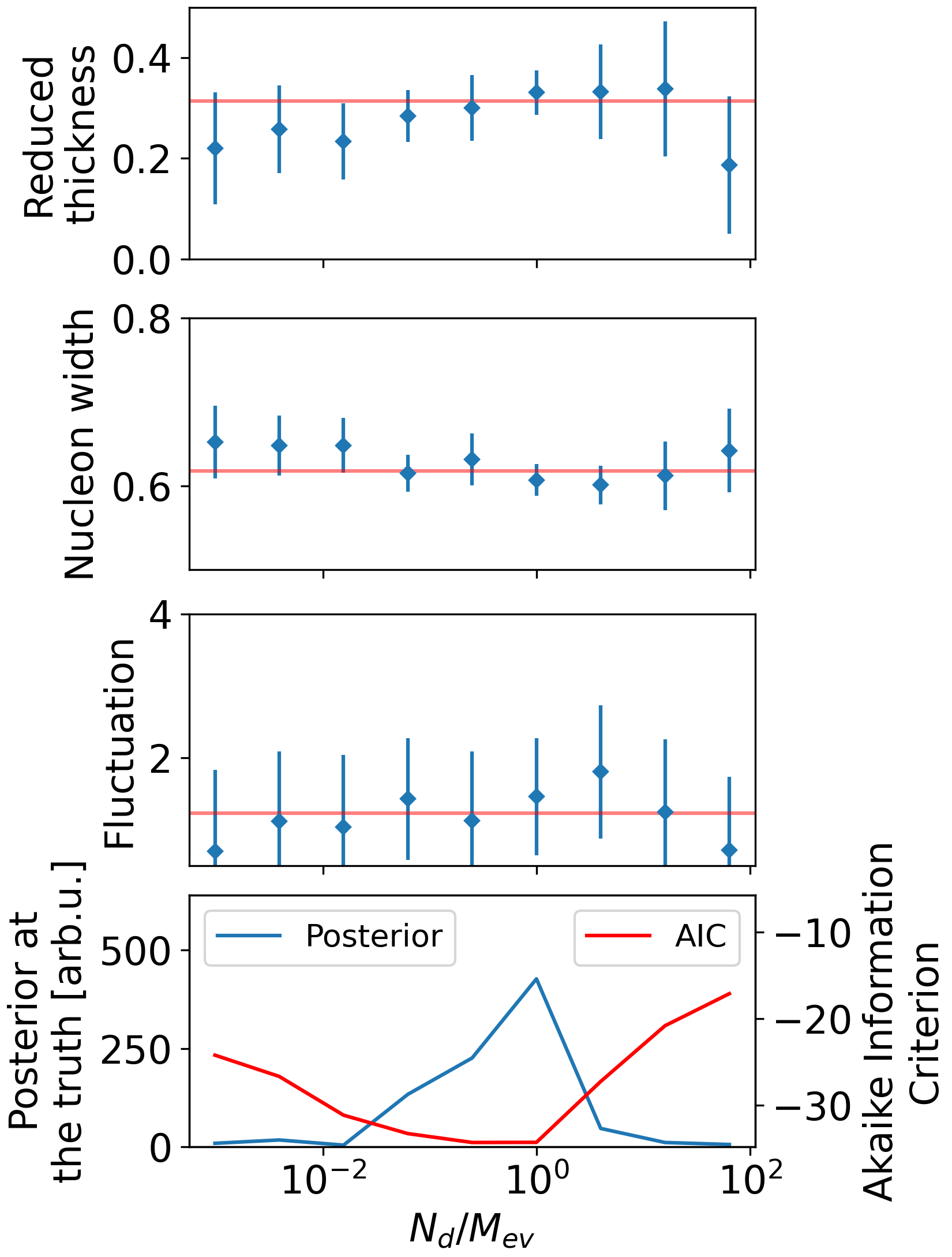}
    \caption{Same as Fig.~\ref{fig:2to16}, with three \trento{} parameters instead of two, and between two (left), three (center) or four (right) observables, with the observables being $\langle \varepsilon_n \rangle$, $n=2,3$, $n=2,3,4$ and $n=2,3,4,5$ respectively.}
    \label{fig:3d}
\end{figure}

Emulator performance depends on multiple factors, in particular the number of parameters, and the complexity of the parameter dependence of the observables.
The results of Bayesian parameter inference evidently also depend on the factors just listed that impact emulator performance, and further depends on the information carried by the different observables.
We repeat the closure test shown on Fig.~\ref{fig:2to16} by adding new observables: we first add $\langle \varepsilon_4 \rangle$, and then $\langle \varepsilon_5 \rangle$.
The results are shown respectively on the left and right panels of Fig.~\ref{fig:e4}. Again, the results remain similar, with the optimal number of design points in the same range as found in previous tests.

We repeat this test, this time with three \trento{} parameters instead of two, starting with two observables ($\langle \varepsilon_n \rangle$, $n=2,3$), then three ($n=2,3,4$) and four ($n=2,3,4,5$) observables. The results are shown respectively on the left, center and right panels of Fig.~\ref{fig:3d}. The results are consistent with all previous ones: closure is successful since the constraints on all parameters are consistent with the truth value of the parameter for any number of design points. Moreover, constraints on the parameters are still best when the ratio of the number of design points $N_d$ to the number of \trento{} simulations per design points $\Mev$ is in the range $N_d/\Mev\approx 0.1$--$1$.
\section{Discussion}

There is a consistent trend among the results shown in Figs.~\ref{fig:2to16}--\ref{fig:3d}: closure was best when the number of design points was slightly smaller than the square root of the total number of events ($\sqrt{N_{\textrm{tot}}}$), which is equivalent to $N_d/\Mev \approx 0.1$--$1$. This trend holds true for all the variations tested in this study.

A rule of thumb that is used at times to determine the number of parameter samples (design points) for a Gaussian process emulator is 10 times the number of parameters ($10 N_{\textrm{params}}$)~\cite{loeppky2009choosing}. This guidance is evidently not meant to be precise, if only because it does not take into account the unavoidable trade-offs between statistical and emulator uncertainty. In the two-parameter case, $N_d = 10 N_{\textrm{params}} \approx 2^4$, which is considerably smaller than the optimal $N_d \approx 2^6$--$2^8$ found in the previous section. We also note that the  optimal number of design points did not change significantly when three model parameters were used instead of two (Fig.~\ref{fig:3d}), which suggests a modest dependence of this optimum on the number of parameters. 

While statistical uncertainties generally converge at the rate $1/\sqrt{\Mev}$,  interpolation uncertainties depend on the details of the model's parameter dependence, and the range of parameters (the prior) over which an emulator is being trained.
Rather than attempting to provide general guidance, we focus on application to heavy ion physics, where our results guidance will apply most directly.

\section{Implications for Bayesian inference in heavy ion collisions}

As seen in the results from Section~\ref{sec:results}, emulator uncertainty is propagated properly in the Bayesian parameter estimation: even if one uses an emulator with large interpolation or statistical uncertainties, the parameter constraints remain consistent with the true value of the parameter used in the closure test. In this sense, we expect the Gaussian process emulators used in Bayesian inference procedure of heavy ion data to be robust. On the other hand, we saw that optimizing the interpolation and statistical uncertainty of the emulator can lead to considerably improved constraints on the model parameters at the same computational cost. In our study, we found an optimal ratio of design points to \trento{} events per design point, $N_d/\Mev$, around $0.1$--$1$.

The optimum $N_d/\Mev \approx 1$ is expected if the interpolation uncertainty in the emulator scales approximately as $1/\sqrt{N_d}$: that is, if both interpolation and statistical uncertainties scale the same, one should aim for a roughly even number for $N_d$ and $\Mev$. Tests published in Ref.~\cite{Nijs:2020roc} found some evidence of $1/\sqrt{N_d}$ convergence of observables in realistic models of heavy ion collisions. 

To apply our work's conclusions to heavy ion collisions, one must remember that there are different sources of stochasticity in the collisions, only one of which was considered in this work: fluctuations in the initial impact geometry of the nuclei. Fluctuations in the number of particles produced will also introduce statistical uncertainty, which was not studied in this work. It is common, though not universal, that the two statistical uncertainties are reduced by oversampling the particles produced for each initial geometry of the collision. The factor $\Mev$ in the ratio $N_d/\Mev \approx 0.1$--$1$ corresponds to the number of initial geometries samples, and \emph{not} the total number of oversamples. 

Finally, we note that the choice of $\Mev$ might be guided by the need for accuracy for certain statistics-hungry observables, rather than the benefits of the emulator as a whole~\cite{Parkkila:2021tqq,Parkkila:2021yha}. It will be useful to repeat the current study with a broader range of observables.
\section{Conclusion}

We studied the interplay between interpolation and statistical uncertainties for Bayesian parameter inference using Gaussian process emulators. We used the \trento{} model of initial conditions in heavy ion collisions, with observables given by the event-averaged spatial anisotropies $\langle \varepsilon_n \rangle$ (Eq.~\ref{eq:averaged_epsilon_n}), which are analogous to momentum anisotropy observables encountered in heavy ion collisions. Using $N_d$ samples of the \trento{} parameter space and $\Mev$ \trento{} events at each parameter point, we performed closure tests to determine the impact of the emulator uncertainty on constraining model parameters. We found that, for a budget of $N_{\textrm{tot}}$ total \trento{} events, the optimal number of design points was around $N_d \approx (0.25$--$1.0) \sqrt{N_{\textrm{tot}}}$, which corresponds to $N_d/\Mev \approx 0.1$--$1$. We found this optimum by looking at multiple different metrics: the maximum value and width of the marginalized posterior, the value of the posterior at the true value of the parameter, and the Akaike Information Criterion (AIC). We found the posterior at the truth to be very effective at quantifying closure. 

In view of our results, we believe it would be beneficial for Bayesian inference applications in heavy ion physics to experiment with different ratios $N_d/\Mev$. Simulating large number of collision events, which reduce all sources of stochastic uncertainties, can give access to statistics-hungry observables~\cite{Parkkila:2021tqq,Parkkila:2021yha}; yet using a large number $\Mev$ of events at each parameter sample will not provide significant benefits for the large number of observables whose emulation is limited by interpolation uncertainty. An improved balance between interpolation and statistical uncertainties can help reduce the emulator's uncertainty below current experimental uncertainties, allowing us to make full use of heavy ion measurements from the RHIC and the LHC.



\section{Acknowledgements}

The authors thank Simon Mak for insightful discussions, and Ron Solz and Matthew Luzum for valuable feedback. B.W. thanks Duke University and Wayne State University for financial support. S.A.B. and J.-F.P. have been supported by the U.S.
Department of Energy Grant no. DE-FG02-05ER41367
This research used resources of the National Energy Research
Scientific Computing Center, a DOE Office of Science User Facility
supported by the Office of Science of the U.S. Department of Energy
under Contract No. DE-AC02-05CH11231 using NERSC award
NP-ERCAP0022230.
Calculations were also performed on the  Duke Compute Cluster.


\bibliographystyle{ieeetr}
\bibliography{biblio_other, biblio_inspirehep}

\begin{thebibliography}{10}

\bibitem{Bass:2001gb}
S.~A. Bass, ``{Microscopic reaction dynamics at SPS and RHIC},'' {\em Nucl.
  Phys. A}, vol.~698, pp.~164--170, 2002.

\bibitem{Gale:2013da}
C.~Gale, S.~Jeon, and B.~Schenke, ``{Hydrodynamic Modeling of Heavy-Ion
  Collisions},'' {\em Int. J. Mod. Phys. A}, vol.~28, p.~1340011, 2013.

\bibitem{Heinz:2013th}
U.~Heinz and R.~Snellings, ``{Collective flow and viscosity in relativistic
  heavy-ion collisions},'' {\em Ann. Rev. Nucl. Part. Sci.}, vol.~63,
  pp.~123--151, 2013.

\bibitem{DerradideSouza:2015kpt}
R.~Derradi~de Souza, T.~Koide, and T.~Kodama, ``{Hydrodynamic Approaches in
  Relativistic Heavy Ion Reactions},'' {\em Prog. Part. Nucl. Phys.}, vol.~86,
  pp.~35--85, 2016.

\bibitem{Philipsen:2012nu}
O.~Philipsen, ``{The QCD equation of state from the lattice},'' {\em Prog.
  Part. Nucl. Phys.}, vol.~70, pp.~55--107, 2013.

\bibitem{Petersen:2010zt}
H.~Petersen, C.~Coleman-Smith, S.~A. Bass, and R.~Wolpert, ``{Constraining the
  initial state granularity with bulk observables in Au+Au collisions at
  $\sqrt{s_{\rm NN}}=200$ GeV},'' {\em J. Phys. G}, vol.~38, p.~045102, 2011.

\bibitem{Novak:2013bqa}
J.~Novak, K.~Novak, S.~Pratt, J.~Vredevoogd, C.~Coleman-Smith, and R.~Wolpert,
  ``{Determining Fundamental Properties of Matter Created in Ultrarelativistic
  Heavy-Ion Collisions},'' {\em Phys. Rev. C}, vol.~89, no.~3, p.~034917, 2014.

\bibitem{Sangaline:2015isa}
E.~Sangaline and S.~Pratt, ``{Toward a deeper understanding of how experiments
  constrain the underlying physics of heavy-ion collisions},'' {\em Phys. Rev.
  C}, vol.~93, no.~2, p.~024908, 2016.

\bibitem{Pratt:2015zsa}
S.~Pratt, E.~Sangaline, P.~Sorensen, and H.~Wang, ``{Constraining the Eq. of
  State of Super-Hadronic Matter from Heavy-Ion Collisions},'' {\em Phys. Rev.
  Lett.}, vol.~114, p.~202301, 2015.

\bibitem{Bernhard:2015hxa}
J.~E. Bernhard, P.~W. Marcy, C.~E. Coleman-Smith, S.~Huzurbazar, R.~L. Wolpert,
  and S.~A. Bass, ``{Quantifying properties of hot and dense QCD matter through
  systematic model-to-data comparison},'' {\em Phys. Rev. C}, vol.~91, no.~5,
  p.~054910, 2015.

\bibitem{Bernhard:2016tnd}
J.~E. Bernhard, J.~S. Moreland, S.~A. Bass, J.~Liu, and U.~Heinz, ``{Applying
  Bayesian parameter estimation to relativistic heavy-ion collisions:
  simultaneous characterization of the initial state and quark-gluon plasma
  medium},'' {\em Phys. Rev. C}, vol.~94, no.~2, p.~024907, 2016.

\bibitem{Moreland:2018gsh}
J.~S. Moreland, J.~E. Bernhard, and S.~A. Bass, ``{Bayesian calibration of a
  hybrid nuclear collision model using p-Pb and Pb-Pb data at energies
  available at the CERN Large Hadron Collider},'' {\em Phys. Rev. C}, vol.~101,
  no.~2, p.~024911, 2020.

\bibitem{Bernhard:2018hnz}
J.~E. Bernhard, {\em {Bayesian parameter estimation for relativistic heavy-ion
  collisions}}.
\newblock PhD thesis, Duke U., 4 2018.

\bibitem{Bernhard:2019bmu}
J.~E. Bernhard, J.~S. Moreland, and S.~A. Bass, ``{Bayesian estimation of the
  specific shear and bulk viscosity of quark\textendash{}gluon plasma},'' {\em
  Nature Phys.}, vol.~15, no.~11, pp.~1113--1117, 2019.

\bibitem{JETSCAPE:2020mzn}
D.~Everett {\em et~al.}, ``{Multisystem Bayesian constraints on the transport
  coefficients of QCD matter},'' {\em Phys. Rev. C}, vol.~103, no.~5,
  p.~054904, 2021.

\bibitem{Nijs:2020ors}
G.~Nijs, W.~van~der Schee, U.~G\"ursoy, and R.~Snellings, ``{Transverse
  Momentum Differential Global Analysis of Heavy-Ion Collisions},'' {\em Phys.
  Rev. Lett.}, vol.~126, no.~20, p.~202301, 2021.

\bibitem{Nijs:2020roc}
G.~Nijs, W.~van~der Schee, U.~G\"ursoy, and R.~Snellings, ``{Bayesian analysis
  of heavy ion collisions with the heavy ion computational framework
  Trajectum},'' {\em Phys. Rev. C}, vol.~103, no.~5, p.~054909, 2021.

\bibitem{JETSCAPE:2020shq}
D.~Everett {\em et~al.}, ``{Phenomenological constraints on the transport
  properties of QCD matter with data-driven model averaging},'' {\em Phys. Rev.
  Lett.}, vol.~126, no.~24, p.~242301, 2021.

\bibitem{Parkkila:2021tqq}
J.~E. Parkkila, A.~Onnerstad, and D.~J. Kim, ``{Bayesian estimation of the
  specific shear and bulk viscosity of the quark-gluon plasma with additional
  flow harmonic observables},'' {\em Phys. Rev. C}, vol.~104, no.~5, p.~054904,
  2021.

\bibitem{Parkkila:2021yha}
J.~E. Parkkila, A.~Onnerstad, S.~F. Taghavi, C.~Mordasini, A.~Bilandzic,
  M.~Virta, and D.~J. Kim, ``{New constraints for QCD matter from improved
  Bayesian parameter estimation in heavy-ion collisions at LHC},'' {\em Phys.
  Lett. B}, vol.~835, p.~137485, 2022.

\bibitem{Moreland:2014oya}
J.~S. Moreland, J.~E. Bernhard, and S.~A. Bass, ``{Alternative ansatz to
  wounded nucleon and binary collision scaling in high-energy nuclear
  collisions},'' {\em Phys. Rev. C}, vol.~92, no.~1, p.~011901, 2015.

\bibitem{Luzum:2013yya}
M.~Luzum and H.~Petersen, ``{Initial State Fluctuations and Final State
  Correlations in Relativistic Heavy-Ion Collisions},'' {\em J. Phys. G},
  vol.~41, p.~063102, 2014.

\bibitem{Ollitrault:1992bk}
J.-Y. Ollitrault, ``{Anisotropy as a signature of transverse collective
  flow},'' {\em Phys. Rev. D}, vol.~46, pp.~229--245, 1992.

\bibitem{Miller:2003kd}
M.~Miller and R.~Snellings, ``{Eccentricity fluctuations and its possible
  effect on elliptic flow measurements},'' 12 2003.

\bibitem{PHOBOS:2006dbo}
B.~Alver {\em et~al.}, ``{System size, energy, pseudorapidity, and centrality
  dependence of elliptic flow},'' {\em Phys. Rev. Lett.}, vol.~98, p.~242302,
  2007.

\bibitem{Alver:2010gr}
B.~Alver and G.~Roland, ``{Collision geometry fluctuations and triangular flow
  in heavy-ion collisions},'' {\em Phys. Rev. C}, vol.~81, p.~054905, 2010.
\newblock [Erratum: Phys.Rev.C 82, 039903 (2010)].

\bibitem{Gardim:2011xv}
F.~G. Gardim, F.~Grassi, M.~Luzum, and J.-Y. Ollitrault, ``{Mapping the
  hydrodynamic response to the initial geometry in heavy-ion collisions},''
  {\em Phys. Rev. C}, vol.~85, p.~024908, 2012.

\bibitem{williams2006gaussian}
C.~K. Williams and C.~E. Rasmussen, {\em Gaussian processes for machine
  learning}, vol.~2.
\newblock MIT press Cambridge, MA, 2006.

\bibitem{low_discrepancy}
M.~Roberts.
\newblock
  \url{https://github.com/j-f-paquet/bayesian_parameter_estimation_tutorial/blob/master/parameter_space_sampling/param_sampling.ipynb}.

\bibitem{niederreiter1992random}
H.~Niederreiter, {\em Random number generation and quasi-Monte Carlo methods}.
\newblock SIAM, 1992.

\bibitem{akaike1974new}
H.~Akaike, ``A new look at the statistical model identification,'' {\em IEEE
  transactions on automatic control}, vol.~19, no.~6, pp.~716--723, 1974.

\bibitem{Liddle:2007fy}
A.~R. Liddle, ``{Information criteria for astrophysical model selection},''
  {\em Mon. Not. Roy. Astron. Soc.}, vol.~377, pp.~L74--L78, 2007.

\bibitem{liddle2004many}
A.~R. Liddle, ``How many cosmological parameters,'' {\em Monthly Notices of the
  Royal Astronomical Society}, vol.~351, no.~3, pp.~L49--L53, 2004.

\bibitem{loeppky2009choosing}
J.~L. Loeppky, J.~Sacks, and W.~J. Welch, ``Choosing the sample size of a
  computer experiment: A practical guide,'' {\em Technometrics}, vol.~51,
  no.~4, pp.~366--376, 2009.

\end{thebibliography}

\end{document}